\documentclass[aps,prl,twocolumn,superscriptaddress]{revtex4-2}

\usepackage{graphicx}
\usepackage{dcolumn}
\usepackage{bm}
\usepackage{amsmath}
\usepackage{amssymb}
\usepackage{latexsym}
\usepackage{epsfig}
\usepackage{amsbsy}
\usepackage{array}
\usepackage{amssymb}
\usepackage{setspace}
\usepackage{color}
\usepackage{url}
\usepackage{hyperref}

\begin{document}
\preprint{Physical Review Letters}
\title{All-optical ultrafast spin rotation for relativistic charged particle beams}

\author{Wen-Qing Wei} \thanks{These authors have contributed equally to this work.} \affiliation{Ministry of Education Key Laboratory for Nonequilibrium Synthesis and Modulation of Condensed Matter, Shaanxi Province Key Laboratory of Quantum Information and Quantum Optoelectronic Devices, School of Physics, Xi'an Jiaotong University, Xi'an 710049, China}	

\author{Feng Wan} \thanks{These authors have contributed equally to this work.} \affiliation{Ministry of Education Key Laboratory for Nonequilibrium Synthesis and Modulation of Condensed Matter, Shaanxi Province Key Laboratory of Quantum Information and Quantum Optoelectronic Devices, School of Physics, Xi'an Jiaotong University, Xi'an 710049, China}

\author{Yousef I. Salamin}  \affiliation{Department of Physics, American University of Sharjah, POB 26666 Sharjah, United Arab Emirates}

\author{Jie-Ru Ren} \affiliation{Ministry of Education Key Laboratory for Nonequilibrium Synthesis and Modulation of Condensed Matter, Shaanxi Province Key Laboratory of Quantum Information and Quantum Optoelectronic Devices, School of Physics, Xi'an Jiaotong University, Xi'an 710049, China}	

\author{Karen Z. Hatsagortsyan} \affiliation{Max-Planck-Institut f\"{u}r Kernphysik, Saupfercheckweg 1, 69117 Heidelberg, Germany}

\author{Christoph H. Keitel} \affiliation{Max-Planck-Institut f\"{u}r Kernphysik, Saupfercheckweg 1, 69117 Heidelberg, Germany}

\author{Jian-Xing Li} \email{jianxing@xjtu.edu.cn} \affiliation{Ministry of Education Key Laboratory for Nonequilibrium Synthesis and Modulation of Condensed Matter, Shaanxi Province Key Laboratory of Quantum Information and Quantum Optoelectronic Devices, School of Physics, Xi'an Jiaotong University, Xi'an 710049, China}	

\author{Yong-Tao Zhao} \email{zhaoyongtao@xjtu.edu.cn} \affiliation{Ministry of Education Key Laboratory for Nonequilibrium Synthesis and Modulation of Condensed Matter, Shaanxi Province Key Laboratory of Quantum Information and Quantum Optoelectronic Devices, School of Physics, Xi'an Jiaotong University, Xi'an 710049, China}	

\date{\today}

\begin{abstract}
An all-optical  method of ultrafast spin rotation is put forward to precisely manipulate the  polarization of relativistic charged particle beams of leptons or ions. In particular, laser-driven dense ultrashort beams are manipulated via single-shot interaction with a co-propagating moderate temporally asymmetric (frequency-chirped or subcycle THz)  laser pulse. Using semi-classical numerical simulations, we find that in a temporally asymmetrical laser field, the spin rotation of a particle can be determined from the flexibly controllable phase retardation between its  spin precession and momentum oscillation. An initial polarization of a proton beam can be rotated to any desired orientation
(e.g., from the common transverse to the more useful longitudinal polarization)  with extraordinary precision (better than 1\%) in tens of femtoseconds using a  feasible frequency-chirped laser pulse. Moreover, the beam qualities, in terms of energy and angular divergence, can be significantly improved in the rotation process.  This method has potential applications in various areas involving ultrafast spin manipulation, like laser-plasma, laser-nuclear and high-energy particle physics.

\end{abstract}


\maketitle

Relativistic beams of spin-polarized  charged particles, such as leptons and ions, play a crucial role in exploring the fundamental structures  \cite{Bass2005,Burkardt2009,Aidala2013} and interactions \cite{Yang2017glue,Acardi2016,Severijns2006,
Gericke2020f},
with the advantage of an increased number of measurable observables and unprecedented test precision \cite{MoortgatPick2008,Aschenauer2019}. They can provide direct access to study  new physics beyond the standard model \cite{Androic2018}, characterize the quantum numbers and chiral couplings of new particles \cite{MoortgatPick2008,Daniele2021}, investigate
spin-dependent nuclear interactions \cite{Rosen1967,Tojo2002mea,Allgower2002}, and increase the cross-section and control the angular distribution of the reaction products in nuclear fusion reactions \cite{Temporal2012,Guillaume2008a}. In the famous proton-spin puzzle \cite{Florian2014,Alexandrou2017}, energetic polarized proton beams are applied
to precisely measure the internal spin-flavor structure of the quarks and the gluon spin distribution of the target proton in proton-proton
collisions  \cite{Adamczyk2014,Adamczyk2016,
Abdallah2021,Aidala2017c,Adler2004d}. In these experiments, the longitudinal spin polarization  is generally sensitive to the helicity distributions of the quarks and gluons \cite{Adamczyk2014,Adler2004d}, while the transverse one is sensitive to the transversity distributions of quarks and parton orbital angular momenta \cite{Aidala2013,Adamczyk2016}. Production of longitudinally spin-polarized (LSP) and transversely spin-polarized (TSP) particle beams with a high degree of polarization 
($\gtrsim70\%$ \cite{MoortgatPick2008,Aidala2013}) is, therefore, quite desirable.

In general, production of a relativistic polarized proton beam is accomplished in two steps. First, a low-energy beam is produced from an optically-pumped polarized $\mathrm{H^-}$ ion source  \cite{Zelenski2002} or a polarized atomic beam source  \cite{Hoffstaetter2006}, and then injected into a high-energy accelerator, where it is commonly rotated to the TSP state in order to suppress the depolarization effect. A relativistic TSP electron beam, by comparison, can be obtained from a low-energy one  produced by photoemission from a GaAs-based cathode \cite{Sinclair2007d} and then accelerated by conventional means, or directly employing the Sokolov-Ternov effect in a storage ring  \cite{Sokolov1967}. 
Relativistic LSP  proton and electron beams
are obtained  via spin-polarization manipulation
in conventional spin rotators \cite{Barber1995ls,Filatov2020} and Siberian Snakes \cite{Lee1997spin,Huang2018high} (the latter are employed to avoid the polarization loss  due to imperfections and intrinsic depolarizing spin resonances \cite{Lee1986r,Lee1997spin}), both of which consisting of a sequence of vertical and horizonal arc dipoles and superconducting solenoids (or specially arranged magnets).
The conventional spin rotators
manipulate the polarization adiabatically, and  the manipulation precision is sensitive to the beam energy spread and the perturbing fields around the particle trajectories. They generally take tens of minutes or even hours to switch between TSP and LSP states, and typically run few times per day in order to cancel the remaining potentially undesirable systematic effects \cite{Aidala2013,Filatov2020}.

Rapid advances in ultraintense ultrashort laser technique, with peak intensities reaching the scale of $10^{23}$ W/cm$^2$, pulse durations of tens of femtoseconds and energy fluctuations of about 1\% \cite{Danson2019,Yoon2019a,Yoon2021b}, offer new avenues for polarized beam generation.
Relatively low-cost and efficient laser-driven plasma accelerators with a gradient exceeding 0.1 TeV/m are capable of providing dense tens-of-MeV proton \cite{Higginson2018, McIlvenny2021,Ren2020} and multi-GeV electron beams \cite{Gonsalves2019}. In all-optical setups, leptons can be
transversely polarized in a standing-wave   \cite{Sorbo2017,Sorbo2018,Seipt2018}, elliptically polarized \cite{Li2019ult,Wan2020ul}, or bichromatic laser pulses \cite{Seipt2019,Songhh2019,Chen2019po,Liu2020tr} due to the quantum  radiative spin-flip effect, since the transverse laser field dominates the spin polarization process; while the more useful LSP lepton beams can only be produced  indirectly  via the helicity transfer from circularly polarized $\gamma$ photons in linear \cite{Zhao2021} or nonlinear Breit-Wheeler pair production processes  \cite{Liyanf2020,Xuekun2021}. The latter is generally  pre-produced via Compton scattering \cite{Phuoc2012al} or bremsstrahlung  \cite{Abbott2016}.  High-energy polarized electron \cite{Nie2021,Wen2019p,Wu2020s} and proton beams \cite{Gong2020ene,Jin2020,Li2021pol} can also be produced  through laser-driven wakefield acceleration of pre-polarized low-energy ones,
generated via a photodissociated hydrogen halide gas target \cite{Spiliotis2021,Sofikitis2018,Rakitzis2003}.
Unfortunately, ultrafast spin manipulation (in particular, spin rotation from TSP to LSP state) of these beams is still a great challenge.

\begin{figure}
\centering
\begin{minipage}[b]{0.5\textwidth}
\centering
\includegraphics[width=3.4in]{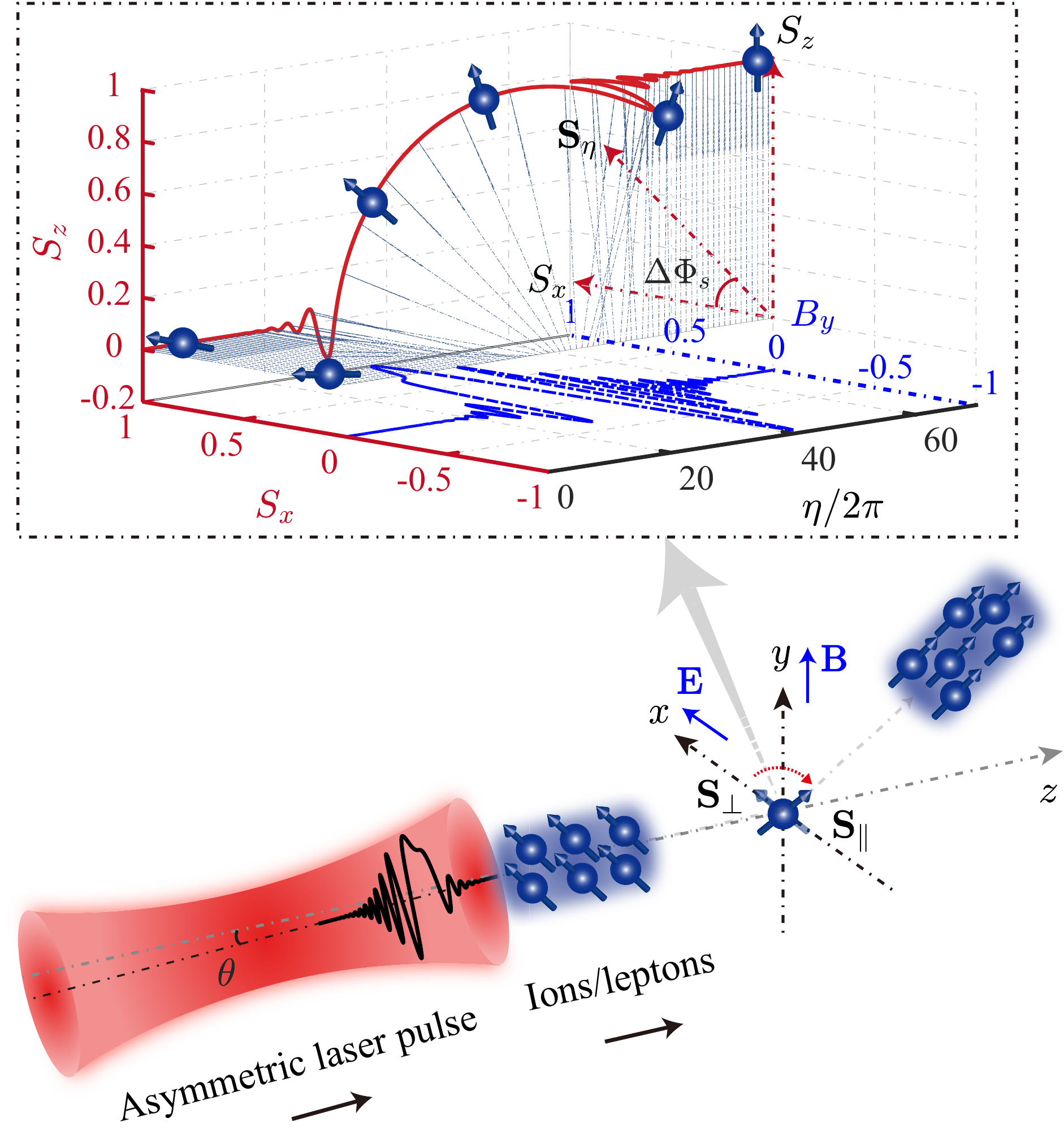}
\end{minipage}
\caption{Interaction scenario. The TSP particle beam  (``$\bf{S}_{\perp}$'', along $+x$ direction and perpendicular to its initial momentum direction) is rotated to the LSP state (``$\bf{S}_{\parallel}$'', parallel to its final momentum direction).    $\bf{E}$ and $\bf{B}$ indicate the electric and magnetic components of the laser field, respectively. (Inset) Particle spin evolution. $\theta$ is the laser incidence angle with respect to the particle beam, $\Delta\Phi_s$ the spin rotation angle  and $\eta$ the laser phase.
  \label{fig1}}
\end{figure}

\begin{figure}
\centering
\begin{minipage}[b]{0.5\textwidth}
\centering
\includegraphics[width=3.4in]{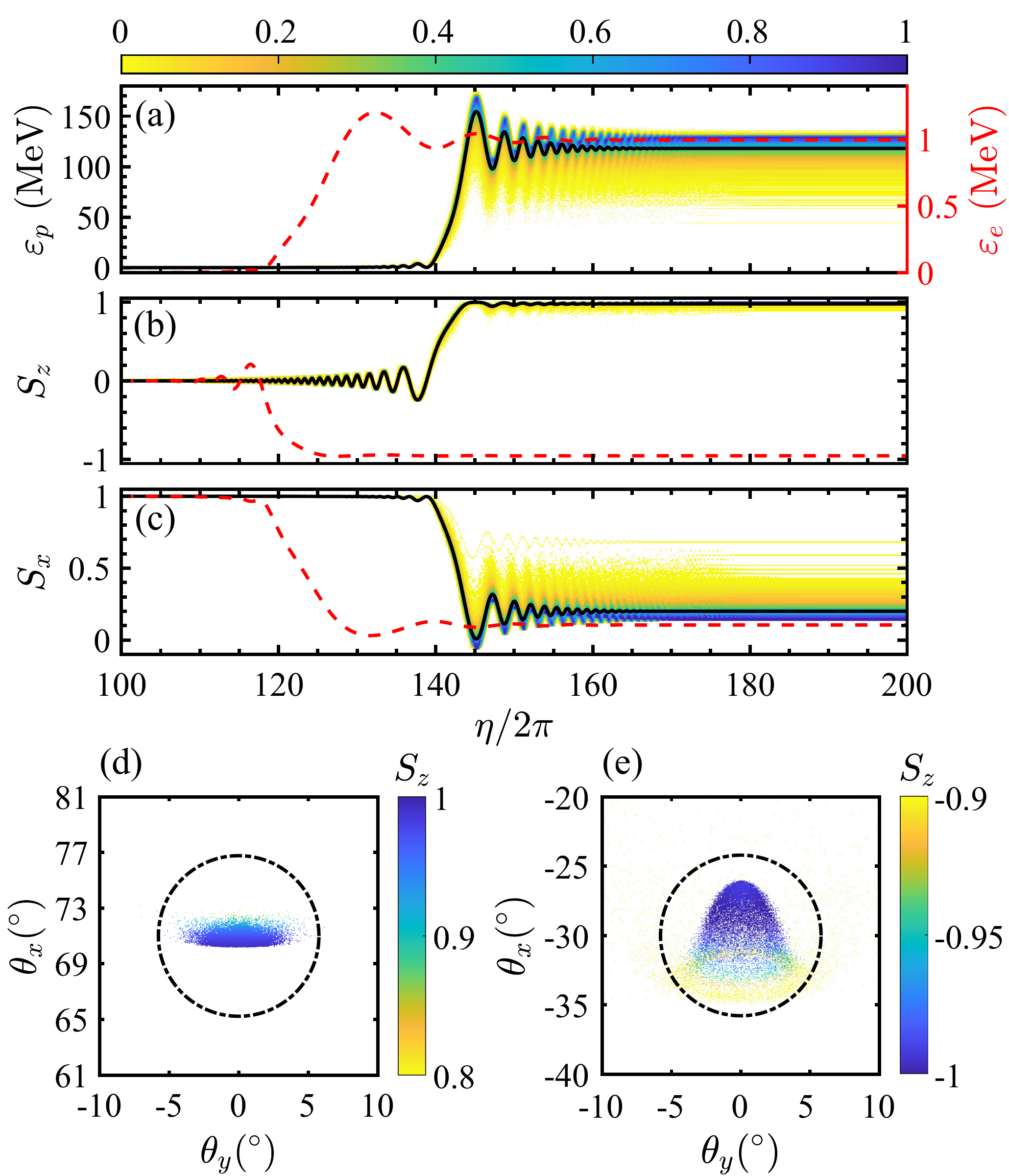}
\end{minipage}
\caption{(a)-(c): Variations of $\mathrm{d}N/\mathrm{d}\varepsilon_p$, $\mathrm{d}N/\mathrm{d}S_z$ and $\mathrm{d}N/\mathrm{d}S_x$ of the proton beam (color) with respect to $\eta$, respectively, and the black curves show the corresponding average values. $N$ is the particle number. The red-dashed curves show the average values for the case of employing an electron beam. (d) and (e): Angle-resolved distributions of $S_z$ vs the transverse deflection angles $\theta_x=\arctan(p_x/p_z)$ and $\theta_y=\arctan(p_y/p_z)$ for the proton and electron beams, respectively. The black-dashed circles denote the corresponding beam initial profile sizes (FWHM).  Simulation parameters are given in the text.\label{fig2}}
\end{figure}

\begin{figure}
\centering
\begin{minipage}[b]{0.5\textwidth}
\centering
\includegraphics[width=3.4in]{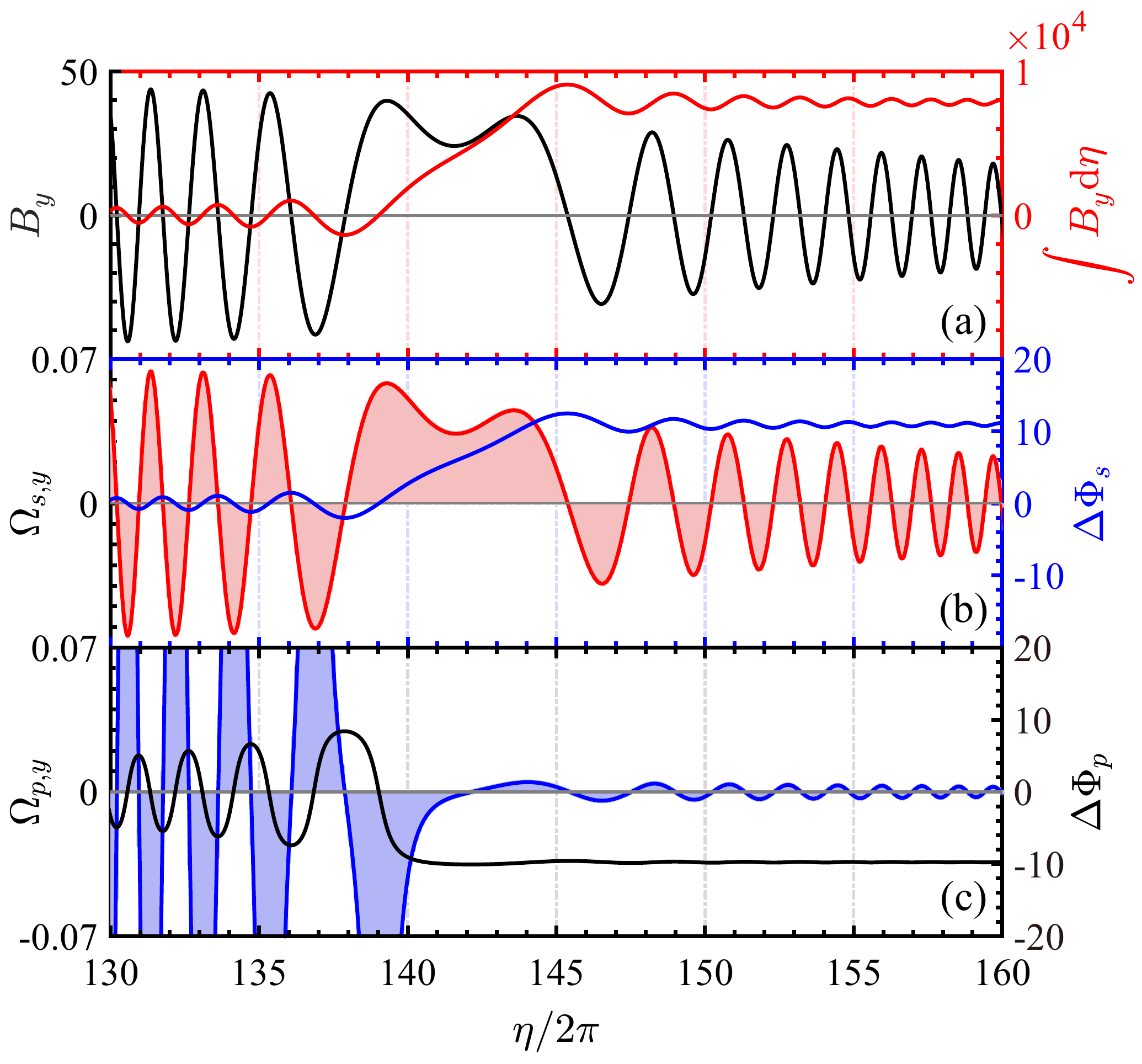}
\end{minipage}
\caption{(a) Evolutions of the normalized magnetic field $B_y$ sensed by the proton (black) and its time-integrated profile (red). (b) Spin precession frequency $\Omega_{s,y}$ (red) and $\Delta\Phi_s$ (blue) vs $\eta$. (c) Momentum angular frequency $\Omega_{p,y}$  (blue) and $\Delta\Phi_p$ (black) vs $\eta$. Other simulation parameters are the same as those in Fig. ~\ref{fig2}. \label{fig3}}
\end{figure}

\begin{figure}
\centering
\begin{minipage}[b]{0.5\textwidth}
\centering
\includegraphics[width=3.4in]{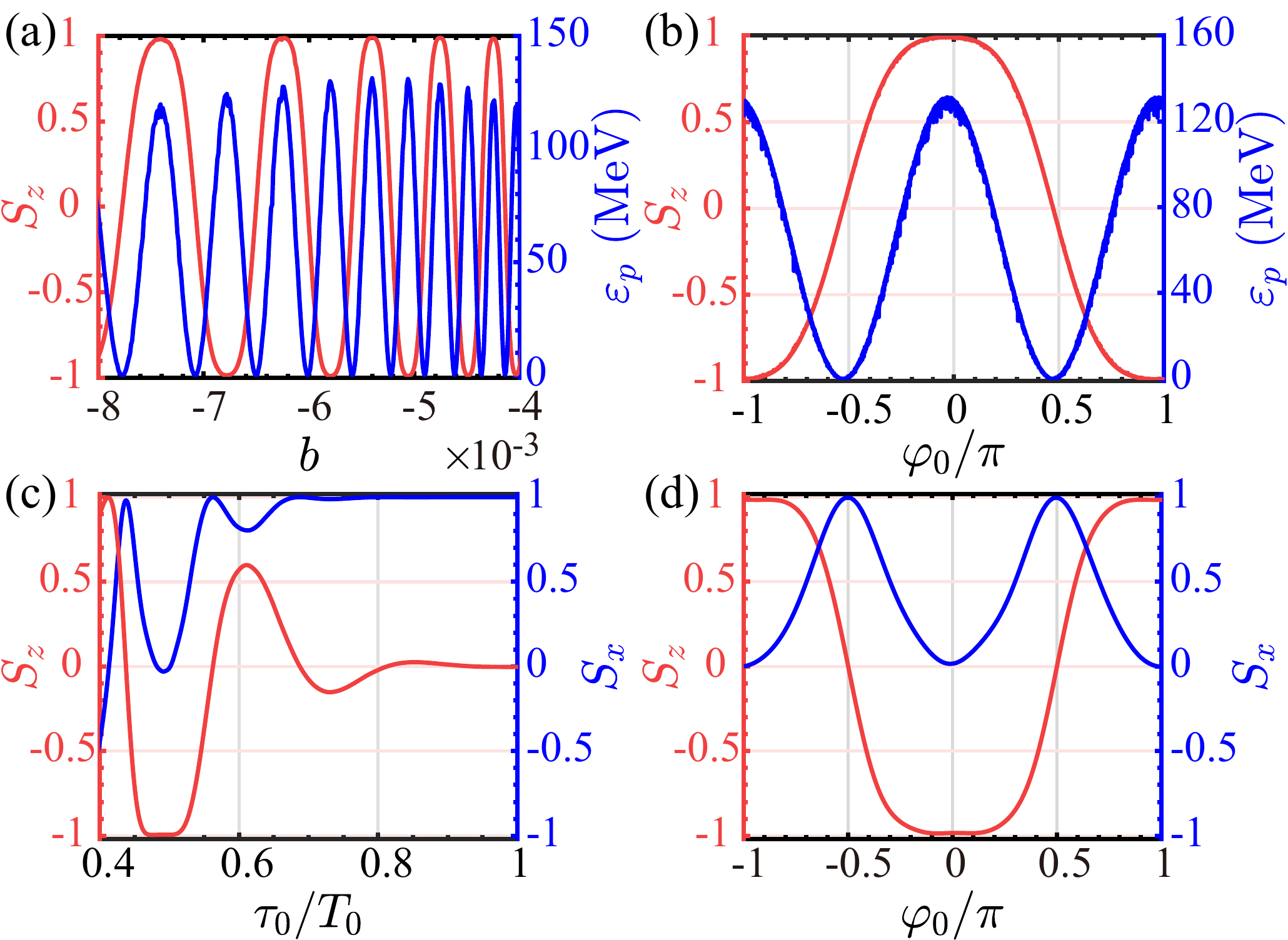}
\end{minipage}
\caption{(a) and (b): Final $S_z$ and $\varepsilon_p$ of the proton beam vs  $b$ and the carrier-envelop phase (CEP) $\varphi_0$ of the chirped laser pulse, respectively.
(c) and (d): Final $S_z$ and  $S_x$ of the electron beam vs  $\tau_0$ and  $\varphi_0$ of the THz  pulse. In (c) and (d), $a_0=2.5$ and $\lambda_0=300~\mu$m. Other parameters are the same as those in Fig.~\ref{fig2}.
\label{fig4}}
\end{figure}

In this Letter, an all-optical ultrafast spin rotation method is put forward to precisely manipulate the polarization of relativistic (laser-driven) ultrashort particle beams through interaction with a co-propagating moderate temporally asymmetrical (e.g., frequency-chirped or sub-cycle THz)  laser pulse (see Fig.~\ref{fig1}).
Focus will be on a TSP  ion or lepton beam to be rotated to the LSP state (or any desired polarization orientation) by a frequency-chirped laser pulse. This will be achieved with  extraordinary precision (better than 1\%  in tens of femtoseconds) while
significantly improving beam qualities in terms of the energy and angular divergence (see Fig.~\ref{fig2}).
We find that in a strong laser field the rotation of a particle's spin is determined from the phase retardation between its  spin precession and momentum oscillation, which can be quite pronounced and flexibly controllable for temporally asymmetrical laser fields (see Fig.~\ref{fig3}). This method 
is demonstrated to be robust with respect to the 
 parameters of the laser and particle beams (see Fig.~\ref{fig4} and \cite{Supplement2021}), is realizable with currently achievable laser facilities, and thus has significant applications in broad areas involving ultrafast spin manipulation.

In our simulations the particle's spin and momentum dynamics  will be treated semiclassically.
For an electron the quantum radiation effects are characterized by the invariant parameter $\chi\equiv|e|\hbar\sqrt{(F_{\mu\nu}p^\nu)^2}/m_e^3c^4\approx a_0\gamma(\hbar \omega_0/m_ec^2)(1-|\bm{\beta}|\cos\theta)$ \cite{Ritus1985,Baier1998}, where $\hbar$ is the  Planck constant and $c$ the speed of light;  $p^\nu$, $\bm{\beta}$, $\gamma$, $e$ and $m_e$ are the 4-momentum, velocity normalized by $c$, Lorentz factor, charge and mass of the electron, respectively. Also, $F_{\mu\nu}$, $\omega_0$, $a_0\equiv|e|E_0/(m_e\omega_0c)$, and $E_0$ are the field tensor, frequency, invariant intensity parameter, and amplitude of the laser pulse, respectively. Note that $\chi$ for a much heavier lepton or ion
is much smaller than that of  an electron in the same  field. With $\chi\ll 1$, the quantum radiation reaction (RR) effects on the particle's momentum and spin (Sokolov-Ternov effect) are negligible
\cite{Baier1967ra,Walser2002,Piazza2012ex,Wen2017s,Nie2021,Wen2019p,Wu2020s,Gong2020ene,
Jin2020,Li2021pol,Gonoskov2021ch}.
Furthermore, the Stern-Gerlach force  is much smaller than the Lorentz force  in our simulations (see Fig.~\ref{fig2}) and can be dropped \cite{Mane2005,Thomas2020}. Thus the particle motion can be adequately described by  Newton-Lorentz equations, and its spin precession is governed by  Thomas-Bargmann-Michel-Telegdi (T-BMT) equation, 
d$\bf{S}$/d($\omega_0 t$) = ${\bf\Omega}_s\times\bf{S}$, with ${\bf\Omega}_s = -\frac{q}{m}[(a+\frac{1}{\gamma}){\bf B}-\frac{a\gamma}{\gamma+1}(\bm{\beta}\cdot{\bf B}){\bm\beta}-(a+\frac{1}{\gamma+1})({\bm\beta}\times{\bf E})]$ \cite{Thomas1926th,Thomas1927p,Bargmann1959}.
Note that we use dimensionless units throughout.  $q$, $m$ and ${\bf\Omega}_s$ are the particle charge, mass and  precession frequency  normalized by $e$, $m_e$ and $\omega_0$, respectively,  $\bf{E}$ and $\bf{B}$ are normalized by $|e|/(m_e\omega_0 c)$, ${\bf S}$ is the particle's spin vector in its rest frame,  and $a=(g-2)/2$ the anomalous magnetic moment of the particle with the gyromagnetic factor $g$. For the proton and electron, $a\approx1.793$ and $1.16\times10^{-3}$, respectively. Quantum
RR effects as
photon recoils and stochastic spin flips must be included for 
$\chi\gtrsim 1$ \cite{Sorbo2017,Sorbo2018,Seipt2018,Li2019ult,
Wan2020ul,Seipt2019,Songhh2019,Chen2019po,
Liu2020tr,Gonoskov2021ch}, while for $10^{-3}\lesssim\chi\ll 1$, RR effects can be treated classically \cite{Piazza2012ex,Guo_2020,Gonoskov2021ch}.

Sample results of spin polarization rotation of proton and electron beams in moderately  frequency-chirped laser pulses are illustrated in Fig.~\ref{fig2}.
An initially TSP particle beam  co-propagating with the laser pulse  has spin $(\bar{S}_x, \bar{S}_y, \bar{S}_z)=(1,0,0)$.  The particles are initially  distributed in a cylinder of radius $R=1.5\lambda_0$ and length $L=0.5\lambda_0$,  and have a number density $n_0\approx2.26\times10^{16}~\mathrm{cm^{-3}}$, and transversely Gaussian and longitudinally uniform profiles. The initial kinetic energy of the proton (electron) beam  is $\varepsilon_p$ = 1 MeV ($\varepsilon_e$ = 0.1 MeV) with 1\% energy spread and angular divergence (FWHM) of about 11.8$^\circ$.
Such pre-polarized beams can be obtained via photodissociation of the aligned hydrogen halide molecules with circularly polarized ultraviolet light \cite{Spiliotis2021,Sofikitis2018,Rakitzis2003}.
We employ a focused, linearly-polarized, and linearly-chirped Gaussia laser pulse, polarized along $+x$ and propagating along $+z$ ($\theta=5^\circ$ in Fig.~\ref{fig1}), with an instantaneous frequency $\omega(\eta)=\omega_0(1+b\eta)$ \cite{Galow2011d,Salamin2012l,Li2012}, where $b$ is the dimensionless chirp parameter, $\eta=\omega_0(t-z/c)$,  and $\omega_0$  the initial frequency at $z=0$. For the proton (electron) beam,
the laser peak intensity is $I_0\approx3.46\times10^{21}~\mathrm{W/cm^2}$ ($5.54\times 10^{16}$ W/cm$^2$) with $a_0\simeq8.5\times10^{-10}\lambda_0[\mathrm{\mu m}]\sqrt{I_0[\mathrm{W/cm^2}]}\simeq50$ (0.2), wavelength $\lambda_0=1~\mu$m, pulse duration $\tau_0=30T_0$ (10$T_0$), period $T_0$, focal radius $w_0=5~\mu$m and $b=-0.00539$ ($-0.0167$).  Such a chirped laser pulse could in principle be  produced in experiments by using dispersion filters \cite{Tournois1997,Verluise2000} (in particular, the linear chirp is adjusted by the group delay dispersion \cite{Weiner2009}) or by the reflection from a relativistic ionization front (interface gas plasma) \cite{Dias1997};
 see also  \cite{Brabec2000,Hajima2003,Gordon2003}. 
For these values, the quantum parameters are $\chi_p\sim 10^{-8}$ ($\chi_e\sim 10^{-7}$) and, hence, the RR effects are negligible in both cases.

Consider what happens as the  front  of the laser pulse catches up and interacts with the protons [in the range of $\eta/2\pi\lesssim 137$ in Figs.~\ref{fig2}(a)-(c)]. Since the front part of the pulse is almost symmetrical [see Fig.~\ref{fig3}(a)], the proton spins oscillate initially only slightly, and  the initial TSP state  is well preserved.  This is followed by interaction with the highly asymmetrical part of the laser pulse, resulting in precise manipulation and rotation of the proton spin from the TSP state ($\bf{S}_\perp$, denoted by $S_x$ in the interaction geometry of Fig.~\ref{fig1}, in which $S_y\approx 0$) to the LSP state ($\bf{S}_\parallel$, denoted by $S_z$) during the range of $137\lesssim\eta/2\pi\lesssim147$. Due to the negative chirp effect, the proton spin polarization can be tuned  in a nearly single elongated cycle [see Fig.~\ref{fig3}(a)]. Finally, the proton interacts with the approximately symmetrical rear part of the laser pulse in the range of $\eta/2\pi\gtrsim147$, and its spin oscillations  become weaker and weaker and eventually die down. The final average longitudinal polarization is $\bar{S}_z\simeq98\%$, symmetrically distributed with respect to $\theta_y$  [see Fig.~\ref{fig2}(d)], and the proton beam is evidently compressed by the laser field (transverse ponderomotive force), particularly in $\theta_x$. 
Moreover, the proton beam is significantly accelerated with an average energy gain of about 120 MeV (that further remarkably decreases the angular divergence $\sim \Delta p_{x,y}/p_z$), due to the phase synchronization between the laser and proton beams \cite{Galow2011d,Sohbatzadeh2009},  with an energy spread $<10\%$ \cite{Supplement2021}.  The energy gain may reach the GeV scale with the appropriate presently available laser parameters \cite{Salamin2012l,Supplement2021}. These results may be optimized further by moderately increasing the laser focal radius, allowing the protons to experience a quasi-uniform laser field.
Such a  highly-LSP low-divergence proton beam with a hundred-MeV energy can be employed as the polarized injection source for high-energy accelerators \cite{Mane2005}, or to search for new physics  by investigating $CP$ or $T-$symmetry \cite{Lenisa2019} (e.g., null high-precision tests of $T$- violation by polarized proton-deuteron scattering at energy of 135 MeV \cite{Przewoski2006}).

For heavier charged  
particles like a deuteron or a triton, the spin rotations and energy gains are moderately reduced compared with those of a proton \cite{Supplement2021}, because the spin manipulation efficiency mainly depends on the charge-to-mass ratio and $g$-factor of the particle \cite{gfactor}. While, for much lighter leptons such as electrons and positrons (with smaller $g$-factors, too), it's vice versa, and consequently a much weaker laser pulse can satisfy the requirements. For instance, 
interacting with a tens-of-GW laser of  $I_0\approx 5.54\times 10^{16}$ W/cm$^2$,   the electron beam can achieve a final polarization of
$\bar{S}_z\simeq-0.95$, with a bell-shaped angular distribution [see Fig.~\ref{fig2}(e)].

The spin manipulation mechanism is analyzed in Fig.~\ref{fig3}.
In the interaction geometry of Fig.~\ref{fig1}, ${\bf E} = \hat{\bm{i}}E_x$ and ${\bf B} = \hat{\bm{j}}B_y$. For protons  with $\gamma\approx1.128$  (see Fig.~\ref{fig2}), the spin precession frequency $|{\bf\Omega}_s|\simeq-\frac{q}{\gamma m}(\gamma a+1)|\bm{\mathrm{B}}|\propto B_y$, yielding $\Omega_{s,x}\simeq \Omega_{s,z}\approx 0$ and $ \Omega_{s,y}\approx 0.064$, according to the T-BMT equation above.
Using a plane-wave laser pulse \cite{Supplement2021}, the spin rotation angle can also be analytically estimated from $\Delta\Phi_s=\int_0^{\eta}|{\bf\Omega}_s|\mathrm{d}\eta'\approx a_0C\sqrt{\frac{\pi}{2b}}[\cos(\frac{1}{4b})\mathrm{F_c}(\frac{1+2b\eta}{\sqrt{2\pi b}})+\sin(\frac{1}{4b})\mathrm{F_s}(\frac{1+2b\eta}{\sqrt{2\pi b}})]+ C_0$,
where $C=(a+1/\gamma)q/m$,  $C_0$ is a constant of  integration,  and $\mathrm{F_c(\cdot)=FresnelC(\cdot)}$ and $\mathrm{F_s(\cdot)=FresnelS(\cdot)}$ are Fresnel cosine and sine integrals. Thus, the proton spin can be expressed as $S_z(\eta)\approx \sin[\theta_0+\Delta\Phi_s(\eta)]$ \cite{Supplement2021} with the initial polarization angle $\theta_0$ (for a TSP  proton beam $\theta_0=0$). $\Delta\Phi_s$ depends sensitively on the time-integrated ${\bf\Omega}_s$  (subsequently on ${\bf B}$) and further on  $a_0$, $\tau_0$, $b$ and  $g-$factor. With the proton momentum direction defined as $\hat{\bm{n}}\equiv\bm{\mathrm{\beta}}/\beta$, its angular frequency  is
${\bf\Omega}_p=(\hat{\bm{n}}\times {\bf E}/\beta-{\bf B})q/\gamma m = [\gamma^2\bm{\beta}\times {\bf E}/(\gamma^2-1)-{\bf B}]q/\gamma m$~\cite{Supplement2021}, which yields  $\Omega_{p,x}\simeq \Omega_{p,z}\approx 0$ and $\Omega_{p,y}\approx 0.5$. Thus, during interaction with the temporally symmetrical laser field, $|{\bf\Omega}_p|\gg|{\bf\Omega}_s|$  with opposite precession directions and the phase of $\Omega_{p,y}$ is ahead of that of $\Omega_{s,y}$ by $\pi$ [see Figs.~\ref{fig3}(b) and (c) in $\eta/2\pi\lesssim 137$ and $\eta/2\pi\gtrsim 147$]. This clearly demonstrates that the proton spin direction  with respect to its momentum direction, i.e., the spin polarization, is unchanged during interaction with the symmetrical part of the laser pulse, where  the spin precession does not result in a net change in $\Delta\Phi_s$ and $\Delta\Phi_p$ simultaneously. Tuning the laser field by an optimal linear negative chirp, its temporal symmetry can be destroyed. This results in an intense quasi-static positive part covering the range $137\lesssim \eta/2\pi\lesssim 147$.  This part exhibits low local instantaneous frequency and small phase variations [see Fig.~\ref{fig3}(a)]. The time-integrated normalized magnetic field is quite large ($\sim8\times10^3$). During interaction with the quasi-static part of the pulse, the positive and negative half-cycles affect the proton momentum and spin dynamics substantially asymmetrically. Meanwhile, the phase retardation  between $\Delta\Phi_s$ and $\Delta\Phi_p$ decreases rapidly and then varies slightly over a relatively long time. Thus a comparable match of integrated spin precession frequency $|\Delta\Phi_s|\simeq6.83\pi/2$  and momentum angular frequency $|\Delta\Phi_p|\simeq6.19\pi/2$  of opposite signs can be acquired with the employed parameters, giving rise to a net phase retardation of about $\pi/2$  [see Figs.~\ref{fig3}(b) and ~\ref{fig3}(c)]. Subsequently, sufficient spin polarization rotation (from TSP to LSP state)  can be achieved. In addition, our numerical simulations demonstrate that spin manipulation of the protons from TSP to LSP state, or vice versa, should satisfy the condition $|\Delta\Phi_s-\Delta\Phi_p|=(2j-1)\pi/2$ [or $\gamma\simeq1+(2j-1)\times0.13$, for given parameters], where $j$ is a positive integer.

For experimental feasibility, impact of other laser and particle beam parameters on the efficiency of the spin rotation is investigated in Fig.~\ref{fig4} and  \cite{Supplement2021}.
Viewed as functions of the chirp parameter $b$ the final $S_z$ and kinetic energy gain $\varepsilon_p$ of the proton beam exhibit local fluctuations of the order of $10^{-3}$, with the optimal rotation occurring, for an optimal $\varepsilon_p$, for $b\approx-0.00539$ [see Fig.~\ref{fig4}(a)] and $\varphi_0=0,\pm\pi$ [see Fig.~\ref{fig4}(b)]. The adjustable precision, $\delta S_z/S_z\sim0.27$,  is estimated using feasible $b$ and $a_0$ parameter values which may result in fluctuations not exceeding $10^{-5}$ \cite{Tournois1997,Verluise2000} and 0.01 \cite{Danson2019,Yoon2019a,Yoon2021b}, respectively.
Note that in experiments the linear chirp parameter $b$  can achieve the required precision
via  mainly adjusting the group delay dispersion, by employing the devices, such as acousto-optic programmable dispersive filters (Dazzler)  \cite{Tournois1997,Verluise2000}, and, the fluctuations in Fig.~\ref{fig4}(a) can be further optimized (slowed down) by simultaneously controlling $\varphi_0$ and moderately enlarging $\tau_0$, which would remarkably reduce the  required precision of $b$.
Also, $\bar{S}_z$  increases exponentially with $a_0$ and decreases with $\tau_0$, while $\bar{S}_x$ follows an opposite trend. The final average energy, $\bar{\varepsilon}_p$, increases almost linearly with $a_0$ and $\tau_0$, while the energy spread remains less than $10\%$ \cite{Supplement2021}.

Besides, the spin rotation process of a realistic subcycle THz  pulse \cite{Schubert2014,Liao2019r,Liao2020t} interacting with a pre-polarized electron beam is investigated here, too, with the fields $B_y \simeq E_x=a_0\cos(\varphi_0+\eta)\exp[-(\eta-4\tau_0/T_0)^2/(\tau_0/T_0)^2]$.  In Figs.~\ref{fig4}(c) and (d),  the optimal LSP state  is obtained for $\tau_0=T_0/2$, which corresponds to
$\varphi_0=0$, $\pm\pi$. The results stay stable within the adjustable range of $\tau_0\pm5\%$ ($\varphi_0\pm9.3\%$) for the parameters employed.  Finally, our  method is widely effective for temporally asymmetrical laser pulses, 
also including the case of
  laser pulses accompanied by additional intense low-frequency magnetic fields \cite{Zhu2015,Williams2020}, however,
 it is not applicable for scenarios that are spatially asymmetrical (e.g., bichromatic) \cite{Supplement2021}.

In conclusion, an all-optical ultrafast spin rotation method has been put forward to precisely manipulate the polarization of relativistic charged particle beams using a  moderate temporally asymmetrical  laser pulse. This method is feasible with currently available laser facilities and would provide a new technique for carrying out polarized beam experiments and high-precision spin-dependent measurements in laser-plasma physics, laser-nuclear physics, and high-energy particle physics.
\\

{\it Acknowledgement}: This work has been supported by the National Natural Science Foundation of China (Grant Nos. U2030104, 11975174, 11775282, 12022506, 11874295 and 12105218), National Key Research and Development Program of China (Grant No. 2019YFA0404900), Chinese Science Challenge Project (Grant No. TZ2016005), and the China Postdoctoral Science Foundation (Grant No. 2020M673367). YIS is supported by an American University of Sharjah Faculty Research Grant (FRG21).






\bibliography{reference}

\end{document}